\documentclass[10pt,journal,compsoc,twoside]{IEEEtran}
\usepackage[utf8]{inputenc}
\usepackage[T1]{fontenc}
\usepackage[pdftex]{graphicx}
\DeclareGraphicsExtensions{.pdf}
\usepackage[nocompress]{cite}
\usepackage[obeyspaces,spaces]{url}
\usepackage{hyperref}
\usepackage{algorithm}
\usepackage[noend]{algorithmic}
\usepackage{array}
\usepackage{amssymb}
\usepackage{mathtools}
\usepackage{interval}
\usepackage{tikz}
\usetikzlibrary{positioning,arrows,fit}
\tikzset{
	switch/.style={
		rectangle,draw,inner sep=0pt,
		minimum height=4mm,minimum width=8mm
	},
	leaf/.style={
		rectangle,draw,fill=gray!70,inner sep=0pt,
		minimum height=4mm,minimum width=8mm
	},
	NIC/.style={
		circle,draw,inner sep=0pt,minimum size=4mm
	},
	>=stealth',
}

\def\assign{\leftarrow}
\def\connabove{\rotatebox[origin=c]{270}{$\Rsh$}}
\def\connbelow{\rotatebox[origin=c]{270}{$\Lsh$}}
\mathchardef\period=\mathcode`.
\DeclareMathSymbol{.}{\mathord}{letters}{"3B} 

\begin{document}

\title{High-Quality Fault Resiliency in Fat-Trees}

\author{
	John~Gliksberg%
	\IEEEauthorrefmark{1}\IEEEauthorrefmark{2}\IEEEauthorrefmark{3},
	Antoine~Capra\IEEEauthorrefmark{2},
	Alexandre~Louvet\IEEEauthorrefmark{2},
	Pedro~Javier~Garc\'\i{}a\IEEEauthorrefmark{3},
	and~Devan~Sohier\IEEEauthorrefmark{1}%
\IEEEcompsocitemizethanks{%
	\IEEEcompsocthanksitem \IEEEauthorrefmark{1} Li-Parad,
	UVSQ, Versailles,
	\{\url{john.gliksberg},\url{devan.sohier}\}\url{@uvsq.fr}
	\IEEEcompsocthanksitem \IEEEauthorrefmark{2} Atos,
	Bruyères-le-Châtel,
	\{\url{antoine.capra},\url{alexandre.louvet}\}\url{@atos.net}
	\IEEEcompsocthanksitem \IEEEauthorrefmark{3} RAAP,
	UCLM, Albacete,
	\url{pedrojavier.garcia@uclm.es}}%
	\thanks{Manuscript received \today ...}%
}

\markboth{IEEE Micro Special Issue on Hot Interconnects}%
{
	Gliksberg \MakeLowercase{\textit{et al.}}:
	High-Quality Fault Resiliency in Fat-Trees
}

\IEEEtitleabstractindextext{%
\begin{abstract}
	Coupling regular topologies with optimised routing algorithms
	is key in pushing the performance
	of interconnection networks of supercomputers.
	In this paper we present Dmodc,
	a fast deterministic routing algorithm
	for Parallel Generalised Fat-Trees (PGFTs)
	which minimises congestion risk
	even under massive network degradation caused by equipment failure.
	Dmodc computes forwarding tables with a closed-form arithmetic formula
	by relying on a fast preprocessing phase.
	This allows complete re-routing of networks
	with tens of thousands of nodes in less than a second.
	In turn, this greatly helps centralised fabric management
	react to faults with high-quality routing tables
	and no impact to running applications
	in current and future very large-scale HPC clusters.
\end{abstract}

}

\maketitle

\IEEEdisplaynontitleabstractindextext

\IEEEraisesectionheading{\section{Introduction}\label{sec:introduction}}

\IEEEPARstart{A}{ majority} of current leading network topologies
for High Performance Computing (HPC) clusters
are fat-tree variants.
(The five most powerful clusters of
the June 2019 Top500 list~\cite{top500}
had fat-tree topologies.)
These networks have some form of static routing tables
computed by a centralised routing engine
and uploaded to all switches.

\begin{figure}
	\centering
	\begin{tikzpicture}
		\node[leaf]   (L1-0)                     {};
		\node[leaf]   (L1-1) [right=6mm of L1-0] {};
		\node[leaf]   (L1-2) [right=8mm of L1-1] {};
		\node[leaf]   (L1-3) [right=6mm of L1-2] {};
		\node[leaf]   (L1-4) [right=8mm of L1-3] {};
		\node[leaf]   (L1-5) [right=6mm of L1-4] {};
		\node[switch] (L2-0) [above=5mm of L1-0] {};
		\node[switch] (L2-1) [above=5mm of L1-1] {};
		\node[switch] (L2-2) [above=5mm of L1-2] {};
		\node[switch] (L2-3) [above=5mm of L1-3] {};
		\node[switch] (L2-4) [above=5mm of L1-4] {};
		\node[switch] (L2-5) [above=5mm of L1-5] {};
		\node[switch] (L3-1) [above=7mm of L2-2] {};
		\node[switch] (L3-2) [above=7mm of L2-3] {};
		\node[switch] (L3-0) [left=6mm  of L3-1] {};
		\node[switch] (L3-3) [right=6mm of L3-2] {};
		\node[NIC] (nic0)  [below=2mm of L1-0,xshift=-3mm] {\footnotesize  0} edge (L1-0);
		\node[NIC] (nic1)  [below=2mm of L1-0,xshift=+3mm] {\footnotesize  1} edge (L1-0);
		\node[NIC] (nic2)  [below=2mm of L1-1,xshift=-3mm] {\footnotesize  2} edge (L1-1);
		\node[NIC] (nic3)  [below=2mm of L1-1,xshift=+3mm] {\footnotesize  3} edge (L1-1);
		\node[NIC] (nic4)  [below=2mm of L1-2,xshift=-3mm] {\footnotesize  4} edge (L1-2);
		\node[NIC] (nic5)  [below=2mm of L1-2,xshift=+3mm] {\footnotesize  5} edge (L1-2);
		\node[NIC] (nic6)  [below=2mm of L1-3,xshift=-3mm] {\footnotesize  6} edge (L1-3);
		\node[NIC] (nic7)  [below=2mm of L1-3,xshift=+3mm] {\footnotesize  7} edge (L1-3);
		\node[NIC] (nic8)  [below=2mm of L1-4,xshift=-3mm] {\footnotesize  8} edge (L1-4);
		\node[NIC] (nic9)  [below=2mm of L1-4,xshift=+3mm] {\footnotesize  9} edge (L1-4);
		\node[NIC] (nic10) [below=2mm of L1-5,xshift=-3mm] {\footnotesize 10} edge (L1-5);
		\node[NIC] (nic11) [below=2mm of L1-5,xshift=+3mm] {\footnotesize 11} edge (L1-5);
		\draw[-,double distance=0.5mm] (L1-0) to (L2-0) to (L1-1) to (L2-1) to (L1-0);
		\draw[-,double distance=0.5mm] (L1-2) to (L2-2) to (L1-3) to (L2-3) to (L1-2);
		\draw[-,double distance=0.5mm] (L1-4) to (L2-4) to (L1-5) to (L2-5) to (L1-4);
		\draw[-] (L3-0) to (L2-0) to (L3-2) to (L2-2) to (L3-0) to (L2-4) to (L3-2);
		\draw[-] (L3-1) to (L2-1) to (L3-3) to (L2-3) to (L3-1) to (L2-5) to (L3-3);
	\end{tikzpicture}
	\caption{$PGFT(3;2.2.3;1.2.2;1.2.1)$
		with leaf switches shown in grey.}%
	\label{fig:pgft}
\end{figure}
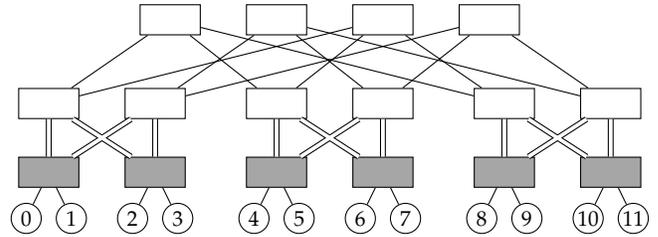

It is sufficient for fat-tree-specific routing algorithms
to be minimal to guarantee deadlock-free routing,
and the regular nature of their target topology class
should simplify load-balancing strategies.
PGFTs~\cite{zahavi2010d} describe all regular fat-trees
for which there is at most one downward switch-path
from any switch to any node
(as shown in Figure~\ref{fig:pgft}).
In this article we refer to fat-trees as PGFTs.
The oblivious routing algorithm for non-degraded PGFTs
(Dmodk~\cite{zahavi2010d}, see Section~\ref{sec:dmodk})
uses this property and their connection logic
to provide load balance through an arithmetic rule.

Due to the sheer amount of equipment
in current and future supercomputers,
hardware failures are to be expected~\cite{domke2014fail}
(especially in optical links~\cite{connors2019simulation}
typically used in higher levels of fat-trees)
and should not hinder running applications as far as possible.
The fabric manager can react to equipment failures
that do not break graph connectivity
by uploading updated routing tables.
In order to do this it requires a fault-resilient routing algorithm
capable of rapid re-routing.
The challenge is to provide these characteristics
while maintaining high-quality static load balance.

Some of the research regarding oblivious fault-resilient routing
focuses on techniques that explicitly target degradations
to regular fat-trees~\cite{zahavi2014quasi}\cite{quintin2016transitively};
there are several re-routing strategies for these techniques.
OpenSM's UPDN~\cite{mlx2007updndoc} 
and Ftree~\cite{zahavi2010optimized} routing engines
can also be applied from scratch to a degraded fat-tree.
PQFT~\cite{zahavi2014quasi} is similar,
though it requires a complete list of faults.
The combination of Dmodk + Ftrnd\_diff~\cite{vigneras2016bxi}
available in BXI~FM~\cite{bxi}
is applied in an offline/online manner
(with an iterative list of network changes
and an up-to-date view of the network),
the goal being fast reaction to faults
with minimal routing changes.
Fabriscale~\cite{villanueva2015routing} also provides
fast centralised re-routing of fat-trees,
by precomputing alternative routes.
A short summary of the limits of the existing approaches
is provided in a previous conference paper~\cite{gliksberg2019hoti}.

The approach that we propose to meet that challenge
is to apply the closed-form arithmetic formula of Dmodk
while relaxing the topological constraint.
For that purpose, we compute shortest paths explicitly
rather than relying on an addressing scheme,
and we balance load according to locally propagated information
rather than relying on level-wide constants.
These two goals are addressed together during preprocessing
and will be the focus of this article, whose main contribution
is the detailed algorithm description in Section~\ref{sec:algo}.

\section{Dmodk}\label{sec:dmodk}

The Dmodk routing algorithm and corresponding PGFT topology
are described in detail in~\cite{zahavi2010d}.
The algorithm relies on a criterion (not shown here)
to determine whether a destination $d$ must be routed
within the down ports and, if so, which one.
Otherwise, an arithmetic formula defines
the up port (with index $p$) to select:

$$
	\textstyle
	p = \left\lfloor \,d\ /\,\prod_{k=1}^lw_k \right\rfloor
	\bmod (w_{l+1}p_{l+1})
$$

The level-wide constants (or \emph{arities})
$w_l$ and $p_l$ respectively denote
the numbers of uplinks and of interlinks
of all switches at level $l$.
With this formula,
each destination's routes are coalesced as early as possible,
and routes to different destinations are spread out as much as possible,
thus minimising collisions between independent traffic.
These closed-form steps rely on a given
organisation of addresses of switches
and indexing of their ports.
Dmodk is a very low complexity and perfectly parallel
routing algorithm for PGFTs,
but it is not applicable to degraded PGFTs
or irregular fat-trees.

\section{Dmodc Description}\label{sec:algo}

The idea behind the fault-resilient algorithm that we propose
is to rely on local information
while using the same closed-form arithmetic formula as Dmodk.
The \emph{c} in Dmodc refers to the neighbouring switches
explicitly determined to be \emph{closer} to the destination
among which paths are chosen.
The aim is fast centralised computation
of routing tables for degraded PGFTs,
providing optimal or well-balanced deterministic routes
even under heavy fabric degradation.
The algorithm begins with a preprocessing phase
(that can be multi-threaded)
followed by a parallel computation phase.
Links are assumed to be bi-directional;
notations used in the expressions hereafter
are defined in Table~\ref{tab:notation}.

\begin{table}[h]
\centering
\caption{Notations Used in Expressions.}\label{tab:notation}
	\begin{tabular}{rl}
		$S$         & is the set of switches \\
		$L$         & is the set of leaf switches ($L\subset S$) \\
		$N$         & is the set of nodes \\
		$E$         & is the set of edges \\
		$\lambda_n$ & is the (only) leaf switch connected to node $n$ ($\lambda_n\in L$) \\
		$\connabove,\connbelow$ & respectively denote down and up links,
			according to rank \\
		$G_s$       & is the ordered list of port groups of switch $s$ \\
		$\Omega_g$  & is the switch connected to port group $g$ \\
		$\#$        & denotes cardinality \\[.08in]
		& \emph{Dmodc-specific notations:} \\
		$c_{s,l}$   & is the cost of switch $s$ to leaf switch $l$ \\
		$\Pi_s$     & is the divider of switch $s$ \\
	\end{tabular}
\end{table}

\subsection{Basic Preprocessing}

For ranking, levels and link directions are determined
according to leaf switches being equivalent to the lowest level.
Groups of ports linked to the same switch are prepared
and sorted by globally unique identifier (GUID)
to help with same-destination route coalescing.

\subsection{Cost}\label{sec:cost}

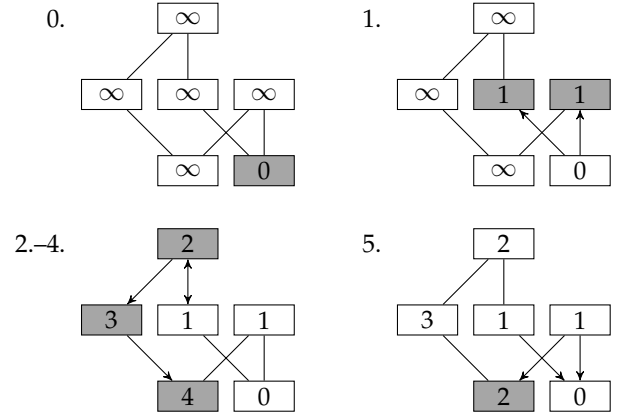
\begin{figure}[h]
	\centering
	\begin{tikzpicture}
		\begin{scope}
			\node[switch] (L1-0)                     {$\infty$};
			\node[leaf]   (L1-1) [right=2mm of L1-0] {0};
			\node[switch] (L2-1) [above=6mm of L1-0] {$\infty$};
			\node[switch] (L2-0) [left=2mm  of L2-1] {$\infty$};
			\node[switch] (L2-2) [right=2mm of L2-1] {$\infty$};
			\node[switch] (L3-0) [above=6mm of L2-1] {$\infty$};
			\draw[-] (L1-1) to (L2-1) to (L3-0) to (L2-0) to (L1-0) to (L2-2) to (L1-1);
			\node (title) [left=11mm of L3-0] {0.};
		\end{scope}
		\begin{scope}[xshift=4.2cm]
			\node[switch] (L1-0)                     {$\infty$};
			\node[switch] (L1-1) [right=2mm of L1-0] {0};
			\node[leaf]   (L2-1) [above=6mm of L1-0] {1};
			\node[switch] (L2-0) [left=2mm  of L2-1] {$\infty$};
			\node[leaf]   (L2-2) [right=2mm of L2-1] {1};
			\node[switch] (L3-0) [above=6mm of L2-1] {$\infty$};
			\draw[-] (L2-2) to (L1-0) to (L2-0) to (L3-0) to (L2-1);
			\draw[->] (L1-1) to (L2-1);
			\draw[->] (L1-1) to (L2-2);
			\node (title) [left=11mm of L3-0] {1.};
		\end{scope}
		\begin{scope}[yshift=-3.0cm]
			\node[leaf]   (L1-0)                     {4};
			\node[switch] (L1-1) [right=2mm of L1-0] {0};
			\node[switch] (L2-1) [above=6mm of L1-0] {1};
			\node[leaf]   (L2-0) [left=2mm  of L2-1] {3};
			\node[switch] (L2-2) [right=2mm of L2-1] {1};
			\node[leaf]   (L3-0) [above=6mm of L2-1] {2};
			\draw[-] (L1-0) to (L2-2) to (L1-1) to (L2-1);
			\draw[<->] (L2-1) to (L3-0);
			\draw[->] (L3-0) to (L2-0);
			\draw[->] (L2-0) to (L1-0);
			\node (title) [left=11mm of L3-0] {2.\---4.};
		\end{scope}
		\begin{scope}[yshift=-3.0cm,xshift=4.2cm]
			\node[leaf]   (L1-0)                     {2};
			\node[switch] (L1-1) [right=2mm of L1-0] {0};
			\node[switch] (L2-1) [above=6mm of L1-0] {1};
			\node[switch] (L2-0) [left=2mm  of L2-1] {3};
			\node[switch] (L2-2) [right=2mm of L2-1] {1};
			\node[switch] (L3-0) [above=6mm of L2-1] {2};
			\draw[-] (L2-1) to (L3-0) to (L2-0) to (L1-0);
			\draw[->] (L2-1) to (L1-1);
			\draw[->] (L2-2) to (L1-0);
			\draw[->] (L2-2) to (L1-1);
			\node (title) [left=11mm of L3-0] {5.};
		\end{scope}
	\end{tikzpicture}
	\caption{Example sequence of cost propagation steps
		in a degraded part of a network.
		Costs to the bottom-right switch are shown in switches.
		At each propagation step,
		the updated costs are in grey.
		Note that in steps~3\---5, some propagations are interrupted
		due to the $c_{s,l}+1<c_{s',l}$ condition in the procedure.
		They could have been achieved with a simple
		$c_{s',l}=\infty$ condition instead;
		however this would have also interrupted
		the propagation of 2 in step~5.
		As a result, the long path on the left
		would not have been avoided.
		For PGFTs (degraded or not),
		such cases are actually impossible
		and the simple condition would suffice;
		but it would not guarantee shortest up\---down paths in
		fat-tree-like topologies.}%
	\label{fig:cost}
\end{figure}

We define the cost $c_{s,l}$
of a switch $s$ to a leaf switch $l$
to be the minimum number of hops between each other
under up\---down restrictions according to rank,
as defined in Procedure~\ref{algo:costdiv}
and illustrated in Figure~\ref{fig:cost}.
This later allows us to determine valid paths
by exploring neighbouring switches and comparing costs.
That exploration could be done here to prepare sets of output ports,
but it's better to leave it for later since each set is only used once
(see Subsection~\ref{sec:dmodc}).
Other all-pairs shortest paths methods
could be substituted here.

\floatname{algorithm}{Procedure}

\begin{algorithm}[H]
\begin{algorithmic}
	\FORALL{$s \in S$}
		\FORALL{$l \in L$}
			\STATE $c_{s,l} \assign \infty$
		\ENDFOR
		\STATE $\Pi_s \assign 1$
	\ENDFOR
	\FORALL{$l \in L$}
		\STATE $c_{l,l} \assign 0$
	\ENDFOR
	\FORALL{$s \in S$ sorted in ascending rank order}
		\STATE $\pi \assign \Pi_s \times
			\#\{s'\ \connabove\ s\}$
		\FORALL{$s'\ \connabove\ s$}
			\FORALL{$l \in L\ |\ c_{s,l}+1 < c_{s',l}$}
				\STATE $c_{s',l}\assign c_{s,l} + 1$
			\ENDFOR
		\ENDFOR
		\FORALL{$s'\ \connabove\ s\ |\ \Pi_{s'} < \pi$}
			\STATE $\Pi_{s'} \assign \pi$
		\ENDFOR
	\ENDFOR
	\FORALL{$s \not\in L$ sorted in descending rank order}
		\FORALL{$s'\ \connbelow\ s$}
			\FORALL{$l \in L\ |\ c_{s,l} + 1 < c_{s',l}$}
				\STATE $c_{s',l}\assign c_{s,l} + 1$
			\ENDFOR
		\ENDFOR
	\ENDFOR
\end{algorithmic}
	\caption{Compute costs and dividers}%
	\label{algo:costdiv}
\end{algorithm}

Thanks to the up\---down restriction,
the complexity of this procedure is in $\mathcal{O}(\#E\#L)$.
This restriction is only for efficiency,
it does not enforce deadlock-freedom.
Some fat-tree-like topologies would result
in up\---down\---up\---down paths
(if such shortcuts appear in neighbouring switches),
since path selection does not distinguish up and down neighbours.
Avoiding this requires a slightly different method:
an extra integer must be stored,
similar to cost but only for downpaths.
More detail can be found in Section~\ref{sec:dmodc}.

In our partially parallel implementation, each worker thread
obtains a block of switches to propagate
with one barrier per level upwards, then downwards.

\subsection{Divider}

\begin{figure}[t]
	\centering
	\begin{tikzpicture}
		\begin{scope}
			\node[switch] (L2-0)                     {3};
			\node[switch] (L2-1) [right=2mm of L2-0] {2};
			\node[switch] (L2-2) [right=2mm of L2-1] {3};
			\node[switch] (L3-0) [above=6mm of L2-0] {1};
			\node[switch] (L3-1) [above=6mm of L2-1] {1};
			\node[switch] (L3-2) [above=6mm of L2-2] {1};
			\draw[-] (L2-2) to (L3-0) to (L2-0) to (L3-1) to (L2-1) to (L3-2) to (L2-2) to (L3-1);
			\node (title) [left=2mm of L3-0] {0.};
		\end{scope}
		\begin{scope}[xshift=4.2cm]
			\node[switch] (L2-0)                     {3};
			\node[switch] (L2-1) [right=2mm of L2-0] {2};
			\node[switch] (L2-2) [right=2mm of L2-1] {3};
			\node[leaf]   (L3-0) [above=6mm of L2-0] {6};
			\node[leaf]   (L3-1) [above=6mm of L2-1] {6};
			\node[switch] (L3-2) [above=6mm of L2-2] {1};
			\draw[-] (L3-0) to (L2-2) to (L3-2) to (L2-1) to (L3-1) to (L2-2);
			\draw[->] (L2-0) to (L3-0);
			\draw[->] (L2-0) to (L3-1);
			\node (div) [below=0mm of L2-0] {\footnotesize $\pi=3\times2$};
			\node (title) [left=2mm of L3-0] {1.};
		\end{scope}
		\begin{scope}[yshift=-2.3cm]
			\node[switch] (L2-0)                     {3};
			\node[switch] (L2-1) [right=2mm of L2-0] {2};
			\node[switch] (L2-2) [right=2mm of L2-1] {3};
			\node[switch] (L3-0) [above=6mm of L2-0] {6};
			\node[switch] (L3-1) [above=6mm of L2-1] {6};
			\node[leaf]   (L3-2) [above=6mm of L2-2] {4};
			\draw[-] (L3-2) to (L2-2) to (L3-0) to (L2-0) to (L3-1) to (L2-2);
			\draw[->] (L2-1) to (L3-1);
			\draw[->] (L2-1) to (L3-2);
			\node (div) [below=0mm of L2-1] {\footnotesize $\pi=2\times2$};
			\node (title) [left=2mm of L3-0] {2.};
		\end{scope}
		\begin{scope}[yshift=-2.3cm,xshift=4.2cm]
			\node[switch] (L2-0)                     {3};
			\node[switch] (L2-1) [right=2mm of L2-0] {2};
			\node[switch] (L2-2) [right=2mm of L2-1] {3};
			\node[leaf]   (L3-0) [above=6mm of L2-0] {9};
			\node[leaf]   (L3-1) [above=6mm of L2-1] {9};
			\node[leaf]   (L3-2) [above=6mm of L2-2] {9};
			\draw[-] (L3-2) to (L2-1) to (L3-1) to (L2-0) to (L3-0);
			\draw[->] (L2-2) to (L3-0);
			\draw[->] (L2-2) to (L3-1);
			\draw[->] (L2-2) to (L3-2);
			\node (div) [below=0mm of L2-2] {\footnotesize $\pi=3\times3$};
			\node (title) [left=2mm of L3-0] {3.};
		\end{scope}
	\end{tikzpicture}
	\caption{Example sequence of divider propagation steps
		in a degraded part of a network.
		Dividers are shown in switches.
		At each propagation step,
		the updated dividers are in grey.
		Note that in step 2,
		the first upswitch is not updated
		because $\pi=2\times2\leq6$.
		Even though there are multiple degradations in the considered case,
		all top switches end up with the divider that they would have had
		in the complete network.}%
	\label{fig:div}
\end{figure}
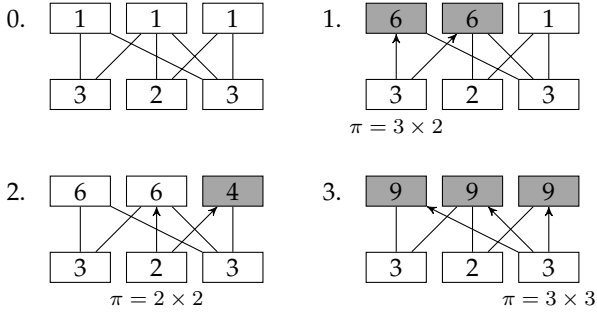

Dmodc is based on the same arithmetic formula as Dmodk.
Prior to the modulo operation,
it begins with an integer division by the product
of $\#\{s'\in S\ |\ s'\ \connabove\ s\}$
(the upward arity of $s$)
of switches at each lower level.
This value represents the number of consecutive destinations
to route through the same port.
It is multiplied when going up levels
to mirror the number of consecutive choices by switches below
before each switch is chosen again.
To reflect the actual state of the network
(in which switches of the same level may have different arities),
only local information must be considered;
in turn, this operation is based
on the products of up-to-date counts of upswitches
(switches connected above),
as defined in Procedure~\ref{algo:costdiv}.
Each downpath corresponds to a potential divider value,
and we choose to keep only the maximum
(as illustrated in Figure~\ref{fig:div}).
The underlying motivation is to generate the same values as in the non-degraded PGFT,
as long as the topological subgroup is not systematically degraded.
The complexity of this part of the procedure is in $\mathcal{O}(\#E)$.

\subsection{Routes Computation}\label{sec:dmodc}

The deterministic output port $p_{s,d}$
and alternative output ports $P_{s,d}$ of every switch $s$
for every destination $d\in N$
(not directly linked to $s$)
are selected with a closed-form formula
based on the results previously determined.
First, port groups leading \textit{closer} to $\lambda_d$
are selected in (\ref{eqn:closer})
(without taking ranking into account),
setting corresponding alternative output ports in (\ref{eqn:dmodc_P}):

\begin{align}
	C_{s,\lambda_d} \assign&\left\{ g \in G_s
	\ |\ c_{\Omega_g,\lambda_d} < c_{s,\lambda_d} \right\}\label{eqn:closer}\\
	P_{s,d} \assign\ &\{p \in g\ |\ g \in C_{s,\lambda_d}\}\label{eqn:dmodc_P}
\end{align}

Selected port groups $C$ are stored in an array
(ordered by GUID of their remote switch),
also represented by $C$:
individual groups are accessed with indices
$i\in\interval[open right]{0}{\#C_{s,\lambda_d}}$
using the $C_{s,\lambda_d}[i]$ notation.
From this, the output port group is chosen in (\ref{eqn:dmodc_pg})
and the port within that group in (\ref{eqn:dmodc_p}):

\begin{align}
	g_{s,d} \assign\ &C_{s,\lambda_d}[\left\lfloor \frac{d}{\Pi_s} \right\rfloor
	\bmod \#C_{s,\lambda_d}]\label{eqn:dmodc_pg} \\
	p_{s,d} \assign\ &
	g_{s,d}[\left\lfloor \frac{d}{\Pi_s\times \#C_{s,\lambda_d}} \right\rfloor
	\bmod \#g_{s,d}]\label{eqn:dmodc_p}
\end{align}

\begin{figure}[t]
\centering
\begin{tikzpicture}
	\begin{scope}
		\node[leaf] (sw) {3};
		\node[switch] (o1) [below=4mm of sw,xshift=-10mm] {4};
		\node[switch] (o2) [above=4mm of sw,xshift=-10mm] {2};
		\node[switch] (o3) [below=4mm of sw,xshift=+10mm] {4};
		\node[switch] (o4) [above=4mm of sw,xshift=+10mm] {2};
		\draw[-] (o3) to (sw) to (o2);
		\draw[-,transform canvas={xshift=+0mm}] (sw) to (o1);
		\draw[-,transform canvas={xshift=+1mm}] (sw) to (o1);
		\draw[-,transform canvas={xshift=-1mm}] (sw) to (o4);
		\draw[-,transform canvas={xshift=+0mm}] (sw) to (o4);
		\draw[-,transform canvas={xshift=+1mm}] (sw) to (o4);
		\node[draw=black,dashed,fit=(o2) (o4)] {};
		\node[left=-0.5mm of o4] {$C_{s,\lambda_{20}}$};
		\node[left=2mm of o2] {(\ref{eqn:closer})};
	\end{scope}
	\begin{scope}[xshift=4.2cm]
		\node[leaf] (sw) {};
		\node[switch] (o1) [below=4mm of sw,xshift=-10mm] {};
		\node[switch] (o2) [above=4mm of sw,xshift=-10mm] {};
		\node[switch] (o3) [below=4mm of sw,xshift=+10mm] {};
		\node[switch] (o4) [above=4mm of sw,xshift=+10mm] {};
		\draw[-] (o3) to (sw);
		\draw[-,dashed] (sw) to (o2);
		\draw[-,transform canvas={xshift=+0mm}] (sw) to (o1);
		\draw[-,transform canvas={xshift=+1mm}] (sw) to (o1);
		\draw[-,transform canvas={xshift=-1mm},dashed] (sw) to (o4);
		\draw[-,transform canvas={xshift=+0mm},dashed] (sw) to (o4);
		\draw[-,transform canvas={xshift=+1mm},dashed] (sw) to (o4);
		\node[left=0mm of o4] {$P_{s,20}$};
		\node[left=2mm of o2] {(\ref{eqn:dmodc_P})};
	\end{scope}
	\begin{scope}[yshift=-2.8cm]
		\node[leaf] (sw) {\footnotesize$\Pi=4$};
		\node[switch] (o1) [below=4mm of sw,xshift=-10mm] {};
		\node[switch] (o2) [above=4mm of sw,xshift=-10mm] {};
		\node[switch] (o3) [below=4mm of sw,xshift=+10mm] {};
		\node[switch] (o4) [above=4mm of sw,xshift=+10mm] {};
		\draw[-] (o3) to (sw) to (o2);
		\draw[-,transform canvas={xshift=+0mm}] (sw) to (o1);
		\draw[-,transform canvas={xshift=+1mm}] (sw) to (o1);
		\draw[-,transform canvas={xshift=-1mm},dashed] (sw) to (o4);
		\draw[-,transform canvas={xshift=+0mm},dashed] (sw) to (o4);
		\draw[-,transform canvas={xshift=+1mm},dashed] (sw) to (o4);
		\node[below=1mm of o4] {$g_{s,20}$};
		\node[left=2mm of o2] {(\ref{eqn:dmodc_pg})};
	\end{scope}
	\begin{scope}[yshift=-2.8cm,xshift=4.2cm]
		\node[leaf] (sw) {};
		\node[switch] (o1) [below=4mm of sw,xshift=-10mm] {};
		\node[switch] (o2) [above=4mm of sw,xshift=-10mm] {};
		\node[switch] (o3) [below=4mm of sw,xshift=+10mm] {};
		\node[switch] (o4) [above=4mm of sw,xshift=+10mm] {};
		\draw[-] (o3) to (sw) to (o2);
		\draw[-,transform canvas={xshift=+0mm}] (sw) to (o1);
		\draw[-,transform canvas={xshift=+1mm}] (sw) to (o1);
		\draw[-,transform canvas={xshift=-1mm}] (sw) to (o4);
		\draw[-,transform canvas={xshift=+0mm}] (sw) to (o4);
		\draw[-,transform canvas={xshift=+1mm},dashed] (sw) to (o4);
		\node[below=1mm of o4] {$p_{s,20}$};
		\node[left=2mm of o2] {(\ref{eqn:dmodc_p})};
	\end{scope}
\end{tikzpicture}
	\caption{
		Example route computation
		with $s$ in grey, $\Pi_s=4$, and $d=20$.
		Costs to $\lambda_{20}$ are shown in switches.
		Indices are ordered from left to right.
		The top-right group is chosen as $g_{s,20}$
		because $\left\lfloor 20/4\right\rfloor\bmod 2=1$,
		and the right port in $g_{s,20}$ is chosen as $p_{s,20}$
		because $\left\lfloor 20/(4\times2)\right\rfloor\bmod 3=2$.
	}\label{fig:dmodc}
\end{figure}
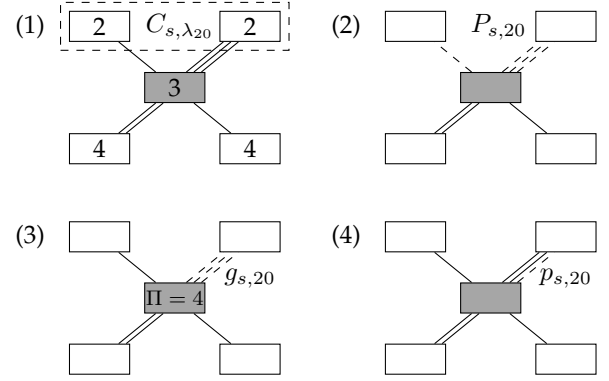

Routes are computed in a loop over leaves
so that $C_{s,\lambda}$ is determined only once
for all nodes connected to $\lambda$
(with $P_{s,d}$ also unchanging $\forall\ d\ |\ \exists\ \lambda_d$).
Figure~\ref{fig:dmodc} illustrates assignments
(\ref{eqn:closer}), (\ref{eqn:dmodc_P}),
(\ref{eqn:dmodc_pg}), and (\ref{eqn:dmodc_p}).

The cost variant for up\---down restriction described in~\ref{sec:cost}
requires (\ref{eqn:closer}) to compare $c$ values for upswitches
and the downpath cost value for downswitches.

\section{Results}\label{sec:results}

The algorithm was implemented in the fabric management software
for Atos's Bull eXascale Interconnect (BXI).
The same code has been used for validation, simulation, and in production.

\subsection{Validity}

Routing is valid for degraded PGFTs
if and only if the cost of every leaf switch
to every other leaf switch is finite:
this reflects every node pair having an up\---down path.
Our implementation includes a pass through all leaf switch pairs
to verify this condition.
The up\---down path restriction is sufficient
to guarantee deadlock-freedom within degraded PGFTs~%
\cite{quintin2016transitively}.

\subsection{Runtime}

\begin{figure}[h]
	\centering
	\includegraphics[width=\linewidth]{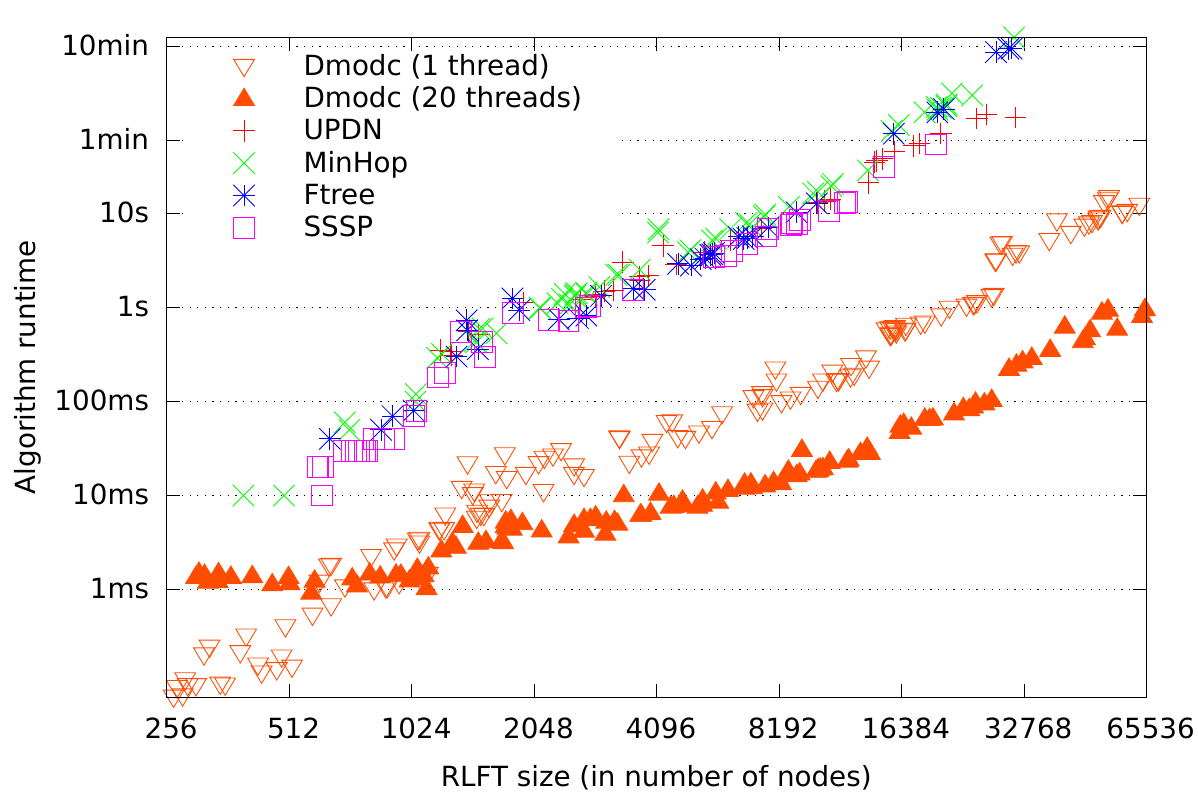}
	\caption{Algorithm runtime
		on a 2.50GHz Intel Xeon E5{-}2680 v3
		for Real-Life Fat-Trees of varying sizes
		(in log\---log scale; lower is better).}%
	\label{fig:times}
\end{figure}

Our C99 implementation had computation of cost, divider, and routes
spread over POSIX threads fetching work with a switch-level granularity.
Figure~\ref{fig:times} reports complete algorithm execution time
alongside OpenSM (version 3.3.21) routing times
(measured by adding timers in the source code)
running on the same machine.
For clusters ranging up to many tens of thousands of nodes,
Dmodc provides fast enough re-routing
for a centralised fabric manager to react to faults
before applications are interrupted.

\subsection{Quality}

The routing algorithm was tested for quality
by generating randomly degraded networks,
computing corresponding routing tables,
and then determining maximum congestion risk
for multiple communication patterns.
This study is available in a previously published
extended abstract~\cite{gliksberg2019hoti}.
The results are comparable or better
than the other available algorithms
across the studied range of degradations.

\section{Conclusion}

The simulation results in Section~\ref{sec:results}
show that Dmodc provides high-quality
centralised fault-resilient routing for PGFTs
at a fraction of the runtime of existing algorithms,
without relying on partial re-routing.
Dmodc is also applicable to fat-tree-like topologies
(as mentioned in Figure~\ref{fig:cost})
but with lower-quality load balancing.
As defined here, no effort has been made to minimise
the size of updates to be uploaded to switches throughout the fabric.

This algorithm is implemented inside BXI~FM~\cite{bxi}
and has been successfully deployed
to an 8490 node PGFT production network
in which it helps provide fault-resiliency
even when faced with thousands of simultaneous changes.

\section*{Acknowledgements}

This research has been undertaken
under a cooperation between CEA and Atos.
with the goal of co-designing
extreme computing solutions.
This research was partly funded by a grant
of Programme des Investissements d'Avenir.
This work has been jointly supported
by the Spanish Ministry of Science, Innovation and Universities
under the project RTI2018{-}098156{-}B{-}C52
and by JCCM under project SBPLY/17/180501/000498.
BXI development was also part of ELCI,
the French FSN (Fond pour la Soci\'et\'e Num\'erique)
cooperative project that associates
academic and industrial partners
to design and provide software components
for new generations of HPC datacenters.

\bibliographystyle{IEEEtran}
\bibliography{dmodc}

\begin{thebibliography}{10}
\providecommand{\url}[1]{#1}
\csname url@samestyle\endcsname
\providecommand{\newblock}{\relax}
\providecommand{\bibinfo}[2]{#2}
\providecommand{\BIBentrySTDinterwordspacing}{\spaceskip=0pt\relax}
\providecommand{\BIBentryALTinterwordstretchfactor}{4}
\providecommand{\BIBentryALTinterwordspacing}{\spaceskip=\fontdimen2\font plus
\BIBentryALTinterwordstretchfactor\fontdimen3\font minus
  \fontdimen4\font\relax}
\providecommand{\BIBforeignlanguage}[2]{{%
\expandafter\ifx\csname l@#1\endcsname\relax
\typeout{** WARNING: IEEEtran.bst: No hyphenation pattern has been}%
\typeout{** loaded for the language `#1'. Using the pattern for}%
\typeout{** the default language instead.}%
\else
\language=\csname l@#1\endcsname
\fi
#2}}
\providecommand{\BIBdecl}{\relax}
\BIBdecl

\bibitem{top500}
E.~Strohmaier, J.~Dongarra, H.~Simon, M.~Meuer, and H.~Meuer, ``Top500,
  https://www.top500.org,'' 1993--2018.

\bibitem{zahavi2010d}
E.~Zahavi, ``D-mod-k routing providing non-blocking traffic for shift
  permutations on real life fat trees,'' \emph{CCIT Report}, 2010.

\bibitem{domke2014fail}
J.~Domke, T.~Hoefler, and S.~Matsuoka, ``Fail-in-place network design:
  interaction between topology, routing algorithm and failures,'' in
  \emph{Proceedings of the International Conference for High Performance
  Computing, Networking, Storage and Analysis (SC)}.\hskip 1em plus 0.5em minus
  0.4em\relax IEEE Press, 2014.

\bibitem{connors2019simulation}
T.~Connors, T.~Groves, T.~Quan, and S.~Hemmert, ``Simulation framework for
  studying optical cable failures in dragonfly topologies,'' in
  \emph{International Parallel and Distributed Processing Symposium
  (IPDPSW)}.\hskip 1em plus 0.5em minus 0.4em\relax IEEE, 2019.

\bibitem{zahavi2014quasi}
E.~Zahavi, I.~Keslassy, and A.~Kolodny, ``Quasi fat trees for hpc clouds and
  their fault-resilient closed-form routing,'' in \emph{22nd Annual Symposium
  on High-Performance Interconnects (HOTI)}.\hskip 1em plus 0.5em minus
  0.4em\relax IEEE, 2014.

\bibitem{quintin2016transitively}
J.-N. Quintin and P.~Vign{\'e}ras, ``Transitively deadlock-free routing
  algorithms,'' in \emph{2nd International Workshop on High-Performance
  Interconnection Networks in the Exascale and Big-Data Era (HiPINEB)}.\hskip
  1em plus 0.5em minus 0.4em\relax IEEE, 2016.

\bibitem{mlx2007updndoc}
OpenSM, \emph{Current OpenSM Routing §~UPDN Routing Algorithm,
  https://github.com/linux-rdma/opensm}, retrieved 2007.

\bibitem{zahavi2010optimized}
E.~Zahavi, G.~Johnson, D.~J. Kerbyson, and M.~Lang, ``Optimized
  {I}nfini{B}and™ fat-tree routing for shift all-to-all communication
  patterns,'' \emph{Concurrency and Computation: Practice and Experience},
  2010.

\bibitem{vigneras2016bxi}
P.~Vign{\'e}ras and J.-N. Quintin, ``The bxi routing architecture for exascale
  supercomputer,'' \emph{The Journal of Supercomputing}, 2016.

\bibitem{bxi}
``Bull exascale interconnect,
  https://atos.net/en/products/high-performance-computing-hpc/bxi-bull-exascale-interconnect.''

\bibitem{villanueva2015routing}
J.~C. Villanueva, T.~Skeie, and S.-A. Reinemo, ``Routing and fault-tolerance
  capabilities of the fabriscale fm compared to opensm,'' Tech. rep. July,
  Tech. Rep., 2015.

\bibitem{gliksberg2019hoti}
J.~Gliksberg, A.~C. Capra, A.~Louvet, P.~J. Garc\'\i{}a, and D.~Sohier,
  ``High-quality fault-resiliency in fat-tree networks (extended abstract),''
  in \emph{26th Symposium on High-Performance Interconnects (HOTI)}.\hskip 1em
  plus 0.5em minus 0.4em\relax IEEE, 2019.

\end{thebibliography}

\begin{IEEEbiographynophoto}{John Gliksberg}
MSc in Computer Science for HPC,
BSc in Mathematics and Physics,
is a PhD student
at Versailles Saint-Quentin-en-Yvelines University (UVSQ), France;
in an international cotutelle with Castilla-La Mancha University (UCLM), Spain;
and in an industrial thesis contract with Atos, France.
His research interests include routing algorithms for HPC interconnects
and he has developed several routing algorithms used in BXI-based clusters.
\end{IEEEbiographynophoto}

\begin{IEEEbiographynophoto}{Antoine Capra}
PhD in Computer Science,
is a lead technical developer in two BXI projects at Atos, France.
His thesis work focused on virtualisation in HPC context.
His research interests include routing algorithms for HPC interconnects
and optimisations to the MPI software layer.
He has contributed to several routing algorithms used in BXI-based clusters.
\end{IEEEbiographynophoto}

\begin{IEEEbiographynophoto}{Alexandre Louvet}
Lead technical developer in multiple BXI projects at Atos, France.
His research interests include
storage, algorithmic, and low-level optimisations for HPC.
He has re-architectured the BXI fabric management software,
developed several routing algorithms used in BXI-based clusters
and improved numerous aspects of BXI's stack.
\end{IEEEbiographynophoto}

\begin{IEEEbiographynophoto}{Pedro Javier Garc\'\i{}a}
PhD in Computer Science,
is an Associate Professor of Computer Architecture and Technology
at Castilla-La Mancha University (UCLM), Spain.
His research focuses on high-performance interconnection networks.
He has published around 60 refereed papers in journals and conferences.
He has guided 2 doctoral theses.
He has coordinated 3 research projects funded
by the Spanish and Castilla-La Mancha Governments.
He has coordinated 4 R\&D agreements between UCLM and different companies.
\end{IEEEbiographynophoto}

\begin{IEEEbiographynophoto}{Devan Sohier}
PhD in Computer Science, is a full professor
at Versailles Saint-Quentin-en-Yvelines University (UVSQ),
within the LI-PaRAD lab.
He has previously been an associate professor at the University of Reims
and at UVSQ.
His primary research interest is distributed algorithms
(in particular probabilistic distributed algorithms in stochastic environments),
with applications to HPC.\@
He has advised 3 PhD students, and is currently advising 2 others.
\end{IEEEbiographynophoto}

\vfill

\end{document}